\documentclass[twocolumn,showpacs,superscriptaddress,preprintnumbers,amsmath,amssymb]{revtex4}

\usepackage{graphicx}
\usepackage{dcolumn}
\usepackage{bm}
\usepackage{color}

\begin{document}

\preprint{}

\title{Dispersion and energy spectrum of spin excitations in an underdoped La$_{1.90}$Sr$_{0.10}$CuO$_{4}$
}

\author{M. Kofu}
\author{T. Yokoo}

\affiliation{
High-Energy Accelerator Research Organization, Tsukuba 305-0801, Japan
}%

\author{F. Trouw}
\affiliation{
Los Alamos National Laboratory, Los Alamos NM 87545, USA
}%

\author{K. Yamada}%
\affiliation{
Institute for Materials Research, Tohoku University, Sendai 980-8577, Japan
}%

\date{\today}

\begin{abstract}

We performed inelastic neutron experiments on underdoped La$_{2-x}$Sr$_{x}$CuO$_{4}$~($x=0.10$, $T_{\rm c}=28.6$~K) using a time-of-flight neutron scattering technique.
Four incommensurate peaks on the two-dimensional reciprocal plane disperse inwards toward an antiferromagnetic zone center as the energy increases.
These peaks merge into a single peak at an energy $E_{\rm cross}$ around  $\hbar\omega =40\pm3$~meV. Beyond $E_{\rm cross}$, the peak starts to broaden and ``hourglass-like'' excitations are observed. 
The  $E_{\rm cross}$  in the underdoped sample is smaller than that reported for the optimally doped La$_{1.84}$Sr$_{0.16}$CuO$_{4}$. The reduction of the $E_{\rm cross}$ is explained by  the doping-independent slope of the downward dispersion below the $E_{\rm cross}$ combined with the smaller incommensurability in the underdoped sample. 
In the energy spectrum of $\chi^{\prime \prime}(\omega)$, we observed a similar "peak-dip-hump" structure in the energy region of  10$\sim45$~meV to that reported for the optimally doped sample.
We discuss the relation between the hourglass-shaped dispersion and the peak-dip-hump energy spectrum. 

\end{abstract}

\pacs{Valid PACS appear here}
\maketitle


One of the central issues regarding the mechanism of high-$T_{\rm c}$ superconductivity is to clarify a common feature of magnetic excitation spectra among high-$T_{\rm c}$ cuprate materials, which has long remained controversial due to the difficulty in inelastic neutron scattering experiment on many different high-$T_{\rm c}$ cuprates.
In recent years, neutron scattering experiments have revealed an overall feature of spin correlations: The spin excitations of hole-doped cuprates form a ``hourglass-shaped'' dispersive branch in energy-momentum space.
In the low-energy region, four symmetric incommensurate~(IC) satellite peaks appear at
($0.5\pm \delta$, 0.5) and (0.5, $0.5\pm \delta$) in the high-temperature tetragonal~(HTT: space group $I4/mmm$) notation~\cite{note1}.
With increasing energy, the IC peaks disperse inwards and merge into a single peak around a characteristic energy, $E_{\rm cross}$. 
As the energy is further increased, the excitations disperse outwards again.
It is noted that in optimally doped YBa$_2$Cu$_3$O$_{6+y}$~(YBCO) a sharp inelastic peak called a resonance peak starts to develop below $T_{\rm c}$ with $\bm Q$=$\bm Q_{\rm AF}$ and $\hbar\omega\sim40$~meV~\cite{Rossat-Mignod1991}, which corresponds to the $E_{\rm cross}$ within the experimental resolution~\cite{P_Bourges2000, D_Reznik2004}.

To date, well-defined hourglass-like dispersive excitations have been observed
in both mono-layered La$_{2-x}$(Sr,Ba)$_x$CuO$_4$~(LSCO, LBCO)~\cite{JM_Tranquada2004_b, NB_Christensen2004, B_Vignolle2007} and 
bi-layered YBCO~\cite{M_Arai1999, P_Bourges2000, SM_Hayden2004, D_Reznik2004, C_Stock2005, V_Hinkov2006},
which indicates that such hourglass-like excitations are common to hole-doped high-$T_{\rm c}$ cuprates. 
However, the microscopic origin of the hourglass-like excitations and their relevance to the high-$T_{c}$ pairing mechanism are still unclear.

Two types of theoretical approaches have been used to understand the hourglass-like excitations. 
One stands on the local spin dynamics in the presence of spin stripes~\cite{CD_Batista2001, GS_Uhrig2004, M_Vojta2006} 
and the other is based on the Fermi liquid theory of itinerant fermions~\cite{I_Eremin2005, MR_Norman2007}.
The former models can describe the overall feature of spin dynamics in a wide energy range in LSCO or LBCO, 
but so far, it has been difficult to reproduce the drastic change in the spin excitation at $T_{\rm c}$ such as the resonance peak in YBCO.
On the other hand, the latter models naturally explain the origin of the resonance peak as a particle-hole bound state, 
but cannot describe quantitatively or even qualitatively the spin excitation above $T_{\rm c}$ or in high energy region well above the resonance.

A recent neutron scattering study on the $\chi^{\prime \prime} (\omega)$ in  La$_{2-x}$Sr$_x$CuO$_4$~(LSCO) with $x=0.16$ ~\cite{B_Vignolle2007}
revealed a two-energy scale or a "peak-dip-hump" structure in the energy spectrum with a peak and a hump located 
at around $\hbar\omega=18$~meV and 45~meV, respectively.
This observation naturally triggered the following questions.
Is the peak-dip-hump structure common for the other cuprates? 
What is the origin of the two-energy scale and the relation with the hourglass-shaped dispersion?
In order to answer to these questions, we performed a neutron scattering experiment on underdoped LSCO with $x=0.10$ 
and observed both the hourglass-like dispersion and the "peak-dip-hump" structure. 
The results obtained by the present study demonstrate not only a common feature 
but also a doping dependence of magnetic excitations of this system. 
We further discuss the different character of magnetic excitations between LSCO and YBCO 
from a view point of duality~(local and itinerant characters) of spins in both systems.


Large single crystals were grown using a traveling-solvent floating-zone~(TSFZ) method, in the same manner as described in Ref.~\cite{M_Kofu2005}.
Three crystals, with each 4~cm long and 8~mm in diameter~(total weight of 40~g) were grown along the [110] direction
in the HTT structure at high temperature.
After the growth, to eliminate oxygen deficiencies, the as-grown crystals were annealed under oxygen gas flow.
We measured magnetic susceptibility under zero-field cooling using a superconducting quantum interference device (SQUID) magnetometer.
The superconducting transition occurs at $T_{\rm c}$(midpoint) of 28.6~K within the transition width $\Delta T_{\rm c}$ of 1.6~K.
The concentrations of La, Sr and Cu ions were precisely determined by inductively coupled plasma (ICP) analysis.

The three crystals were coaligned using a triple-axis spectrometer in JRR-3M.
Inelastic neutron scattering experiments were performed on the chopper spectrometer PHAROS at Los Alamos National Laboratory~(LANL)
using the time-of-flight neutron scattering technique.
The incident beam was parallel to the $c$-axis~($L$-direction) of sample,
and the horizontal plane corresponded to the [$H,H,L$] zone.
Using position sensitive detectors~(PSD), this configuration makes it possible for us to survey peaks around $\bm Q_{\rm AF}=(0.5, 0.5)$ 
in the two dimensional ($H,K$) plane.
Measurements were carried out using three different incident neutron beams characterized by both incident energy $E_i$ and chopper frequency $f$,
$E_i=40$~meV and $f=120$~Hz, $E_i=60$~meV and $f=240$~Hz, and $E_i=80$~meV and $f=300$~Hz.
%


\begin{figure}[tb]
\begin{center}
\includegraphics[width=0.7\hsize]{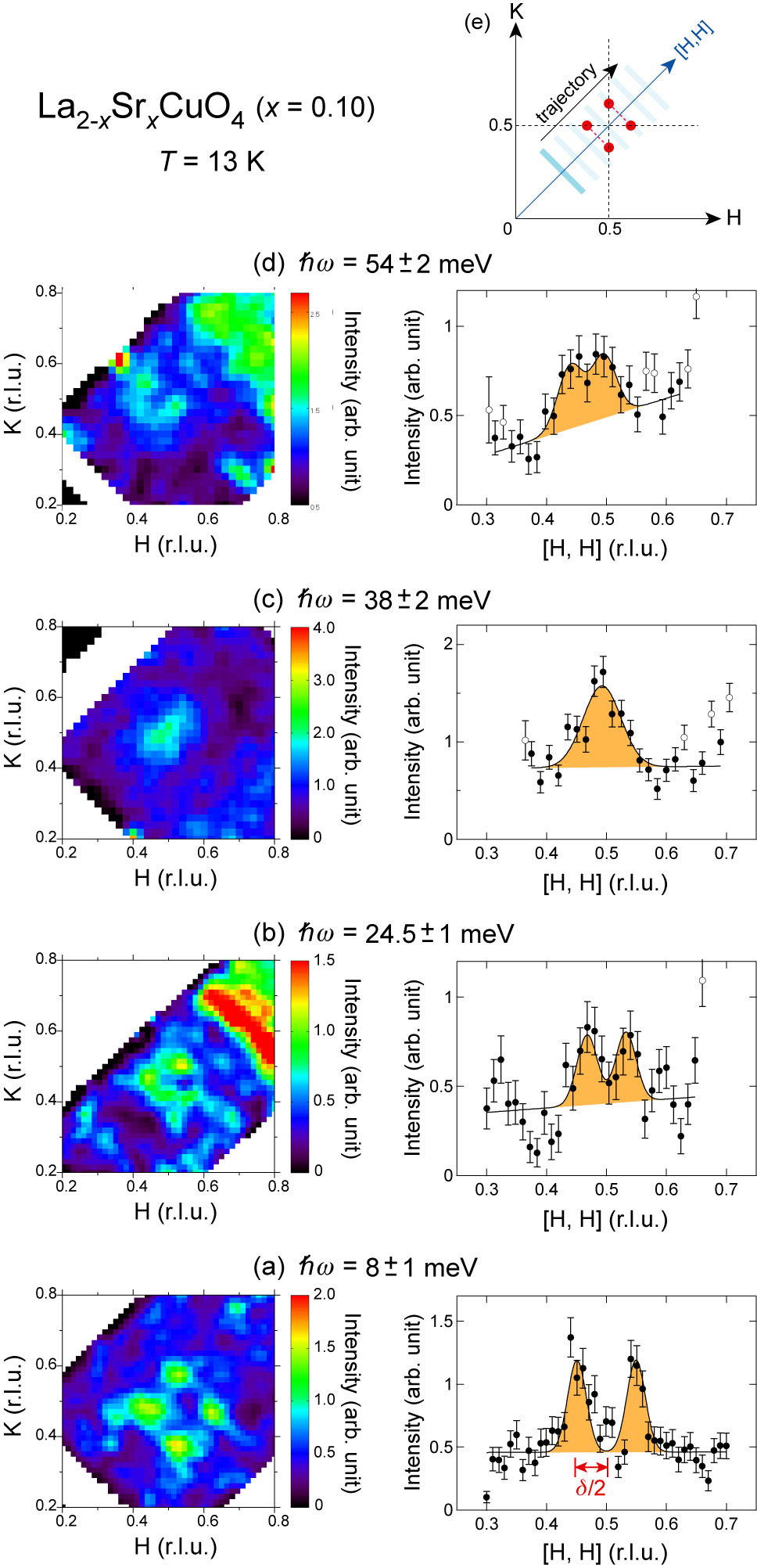}
\caption{
Left figures show constant-energy slices of magnetic excitations of La$_{1.9}$Sr$_{0.1}$CuO$_{4}$
at (a)~$\hbar\omega=8$~meV, (b)~$\hbar\omega=25$~meV, (c)~$\hbar\omega=38$~meV, and (d)~$\hbar\omega=57$~meV.
All data are taken at $T=13$~K.
The data (a)-(b) were measured with an incident energy $E_i=40$~meV and the data (c)-(d) with $E_i=80$~meV.
Right figures show one dimensional $\bm Q$ cuts along the [$H,H$] direction.
The intensities are integrated over the shaded area in (e).
Solid lines are the result of fits with assuming a single or a pair of Gaussian peak and a linear background.
}
\label{fig:map}
\end{center}
\end{figure}

\begin{figure}[tb]
\begin{center}
\includegraphics[width=0.8\hsize]{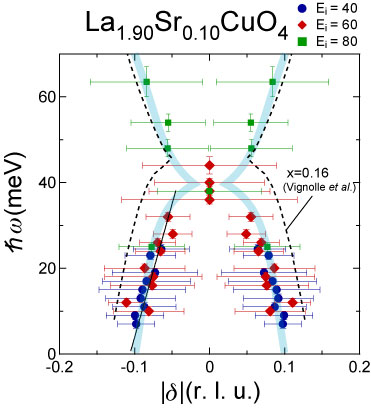}
\caption{
Dispersion relation for La$_{1.9}$Sr$_{0.1}$CuO$_{4}$ at 13~K with incident energies of 40~(blue circles), 60~(red diamonds),
and 80~meV~(green squares).
Dashed line denotes the smoothed dispersion of La$_{1.84}$Sr$_{0.16}$CuO$_{4}$ at 10~K~(Ref.~\cite{B_Vignolle2007}).
In the figure, the obtained values of incommensurability $\pm\delta$ are plotted as a function of energy.
Horizontal bars are not error bars but correspond to the peak-widths (FWHM).
Solid blue line indicates a guide to the eye of dispersion.
}
\label{fig:dispersion}
\end{center}
\end{figure}

Constant-energy slices of magnetic excitations at $T=13~K$ are shown in the left panels of Fig.~\ref{fig:map}.
It is noted that the background of the spectrometer is low and that only weak random noise was subtracted from the raw data.
For each energy transfer, a clear signal can be observed around $\bm Q_{\rm AF}=(0.5, 0.5)$. Particularly, at $\hbar\omega=8$~meV, 
four well-defined incommensurate peaks surround $\bm Q_{\rm AF}$ and the incommensurability, $\delta$, is determined  to be $\sim 0.1$~(r.l.u.),
which is consistent with a triple-axis neutron scattering measurements~\cite{CH_Lee2000}. 
These satellite peaks come closer together with increasing energy transfer, $\hbar\omega$, 
indicating the excitation disperses inwards toward $\bm Q_{\rm AF}$.
We note the arc-like strong scattering around $\bm Q=(0.7,0.7)$ at $\hbar\omega=25$~meV is not an intrinsic signal.
As seen in the figure at $\hbar\omega=38$~meV, a single peak was found at $\bm Q_{\rm AF}=(0.5,0.5)$
and the excitation tends to disperse outwards in both directions to the higher and lower energy regions.
Therefore, the hourglass-like excitations also exist in underdoped LSCO with $x=0.10$.

To analyze the data more quantitatively, we plotted the result of one dimensional scan along the [$H,H$] direction of the data 
in the left figures~(see right figures in Fig.~\ref{fig:map}).
We integrated intensities along the direction perpendicular to [$H,H$], $-0.1 \leq [\bar H,H] \leq 0.1$~r.l.u., 
as shown in the shaded area of Fig.~\ref{fig:map}(e).
By this procedure, four incommensurate peaks at $\bm Q=$~($0.5\pm \delta$, 0.5) and (0.5, $0.5\pm \delta$) are projected 
on ($0.5(1 \pm \delta)$, $0.5(1 \pm \delta)$). 
Hence the observed peak splitting corresponds to the half of $\delta$ as depicted in the right figure in Fig.~\ref{fig:map}(a)).
Solid lines are the results of the fits with assuming a pair of Gaussian peak and a linear background.
We drew a dispersion relation in Fig.~\ref{fig:dispersion} by plotting the observed values of  $\delta$ in both positive and 
negative horizontal axis directions.

In the energy region, $37 \leq \hbar\omega \leq 43$~meV, a single peak was observed 
and the $E_{\rm cross}$ is evaluated to be $40\pm3$~meV, which is smaller than that of the $x=0.16$ sample. 
On the other hand, the slope of the dispersion for the downward excitation~($\hbar\omega < 40$~meV) is almost the same for the two samples 
as seen in Fig.~\ref{fig:dispersion}, which was pointed out by Christensen~{\it et al.}~\cite{NB_Christensen2004}. 
Here, we define the spin wave velocity $\hbar v$ as $dE/d\delta$. 
The solid line in the figure corresponds to $\hbar v_{\rm down}=375$~meV/\AA $^{-1}$.
The smaller value of the $E_{\rm cross}$ for the underdoped sample can be understood by the doping-independent spin wave velocity 
of the downward excitation. 
The $E_{\rm cross}$ will be proportional to the $\delta$ at low energies. 
Particularly in underdoped LSCO, the $E_{\rm cross}$ is expected to depend linearly on both doping rate, $p$, and the maximum $T_{c}$ 
because of the linear relation among $\delta$, $p$ and the maximum $T_{\rm c}$ ~\cite{K_Yamada1998}.
This prediction is supported by the previous result on LSCO with $x=0.07$~\cite{H_Hiraka2001} 
where the doping dependence of magnetic peak-width suggests that the $E_{\rm cross}$ is smaller than that of LSCO with $x=0.10$.

In contrast to the weak doping dependence of the spin wave velocity for the downward excitation, 
the spin wave velocity for the upward excitation is remarkably doping dependent. 
For example, $\hbar v_{\rm up}=510 \pm 50$~meV/\AA $^{-1}$~\cite{B_Vignolle2007} is reported for the $x=0.16$ sample 
which is much smaller than that of undoped La$_2$CuO$_4$, $\hbar v_{\rm SW} = 850 \pm 30$~meV/\AA $^{-1}$~\cite{G_Aeppli1989}.
Such different responses to doping indicate different origins of the downward and the upward excitations. 
Additionally, such different characters of spin excitations are also observed in the peak-widths, 
shown as the horizontal bars in Fig.~\ref{fig:dispersion}; 
for the upward excitation the peak-widths are larger than those for the downward excitation as seen in $x=0.16$~\cite{B_Vignolle2007}.

\begin{figure}[tb]
\begin{center}
\includegraphics[width=0.9\hsize]{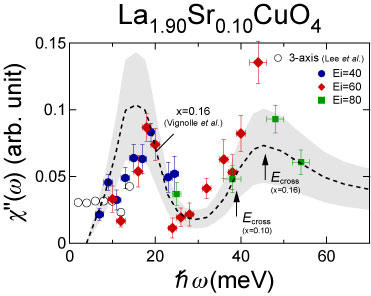}
\caption{
Energy dependence of $\chi^{\prime \prime} (\omega)$ for La$_{1.9}$Sr$_{0.1}$CuO$_{4}$ at 13~K with incident energies of 40~(blue circles),
 60~(red diamonds), and 80~meV~(green squares).
Dashed line shows the results of La$_{1.84}$Sr$_{0.16}$CuO$_{4}$ at 10~K~(Ref.~\cite{B_Vignolle2007}).
Open circles correspond to the data taken at 9~K using triple axis spectrometer~(Ref.~\cite{CH_Lee2000})
Arrows indicate the cross-point of hourglass excitation.
}
\label{fig:chi}
\end{center}
\end{figure}

We next show the energy spectrum of the $Q$-integrated (local) dynamical magnetic susceptibility $\chi^{\prime \prime} (\omega)$ at $T=13$~K.
Here, $\chi^{\prime \prime} (\omega)$ is obtained by correcting for the thermal population factor 
in the integrated dynamical structure factor $S(\omega) = \int S(\bm Q, \omega) d\bm Q$. 
Since the thermal population factor $1/(1-\exp(-\hbar\omega/k_B T)) \simeq 1$ for $\hbar\omega \geq  5$~meV at $T=13$~K, 
$\chi^{\prime \prime} (\omega)$ and $S(\omega)$ are nearly equivalent.
$S(\omega)$ was estimated from the fits of magnetic signals assuming a single or a pair of 
Gaussian peaks~(see in right panels of Fig.~\ref{fig:map}).
In order to draw an energy spectrum over a wide energy range, we normalized each data point of $\chi^{\prime \prime} (\omega)$ 
taken under different incident beams. 
For the normalization we considered the differences of beam flux, energy resolution, and magnetic form factor~\cite{note2} for each data point.

In Fig.~\ref{fig:chi}, we plot the result of $x=0.10$~(our data) in addition to the result from the triple axis scattering 
for $x=0.10$~(Lee {\it et al.}~\cite{CH_Lee2000}).
Amplitudes of $\chi^{\prime \prime} (\omega)$ from two different experiments are scaled using the value at $\hbar\omega=10$~meV~\cite{note3}.
In the figure, we also drew a smoothed band to represent the result for $x=0.16$~(Vignolle {\it et al.}~\cite{B_Vignolle2007}). 
The band width approximately represents the uncertainty in the absolute value of $\chi^{\prime \prime} (\omega)$ 
between the different experimental results.
Interestingly, $\chi^{\prime \prime} (\omega)$ exhibits a similar "peak-dip-hump" structure to that is observed in LSCO with $x=0.16$. 
The peak and hump positions weakly depend on doping.

Here, we shortly remark about the possibility of contamination by phonons into the peak at around 20~meV, 
since the background from phonons are often intense in this energy range. 
In order to distinguish and remove clear contamination by phonons, we analyzed the $q$-profiles of IC peaks at many different energy values 
by using different incident neutron energies. 
If we observed any anomalous profiles possibly due to phonon scattering, we discarded the data point. 
Therefore, the peak around 20~meV is considered not to be severely contaminated by phonons, 
though we are not completely free from the effect from non-dispersive phonons or possible coupling between phonons and magnetic excitations.

As seen in Fig.~\ref{fig:chi}, in the underdoped sample the peak intensity at around 20~meV is relatively weaker than 
that observed in the optimally doped sample. 
Furthermore, the peak at around 20~meV for $x=0.16$ is strongly suppressed at $T=300$~K~\cite{B_Vignolle2007}.
Thus, it is natural to consider that the peak is related to the onset of superconductivity either directly or indirectly.
Here we compare the peak observed in LSCO with the resonance peak in YBCO. In YBCO, the resonance peak position, $E_r$, 
nearly corresponds to that of the $E_{\rm cross}$ and is approximately proportional to $T_{\rm c}$.
On the other hand, in LSCO, as indicated by arrows in Fig.~\ref{fig:chi}, 
the $E_{\rm cross}$ locates near the hump energy which is higher than the peak energy. 
Furthermore, the resonance peak in optimally doped YBCO appears below $T_{\rm c}$ with sharp widths in energy and $Q$, 
while the peak in LSCO is much broader and perhaps remains even above $T_{\rm c}$~\cite{CH_Lee2003}.
Therefore, although the hourglass shape of the magnetic dispersion is common, the energy spectrum and 
its temperature dependence are different between LSCO and YBCO.

Recently, phenomenological theories, which take into account both itinerant fermions and local spins~\cite{MV_Eremin2006, Y_Bang2007} 
have successfully reproduced the experimental excitation data for YBCO in both superconducting and normal states. 
Within this framework, we speculate that the remaining IC correlation well below $E_{\rm cross}$ in LSCO may be related with 
the local spin state such as a stripe state. 
Moreover, the aforementioned distinctions in the magnetic excitations between LSCO and YBCO possibly originate from 
the different local spin state and/or the different degree of dual nature between the two systems, 
more localized nature in LSCO and more itinerant nature in YBCO.

Furthermore, it should be noted that the degree of such duality seems to correlate with the pseudo-gap behavior. 
In LSCO, the doping region of the pseudo-gap extends towards the over-doped region~\cite{CH_Lee2003, Y_Ando2004}, 
while in YBCO the pseudo-gap is mainly seen in the underdoped region. 
Furthermore, in the underdoped YBCO a peak-dip-hump structure was reported by Dai et al~\cite{P_Dai1999} as seen in the optimally doped LSCO. 
The similar peak-dip-hump structure between the optimally doped LSCO and the underdoped YBCO may correspond to such different doping regions 
for the pseudo-gap behavior. 
In other words, the stability of the pseudo-gap depends on the dual nature of magnetic properties in high-$T_{\rm c}$ cuprates.

Finally, we briefly address the low energy component below around 10~meV which appears in the underdoped LSCO. 
We speculate the growth of the intensity below around 10~meV coincides with the growth of static 
or quasi static spin correlations commonly observed in the underdoped region.  
We note that in LSCO with $x=0.07$, the low energy component grows more rapidly than in LSCO with $x=0.1$ 
with decreasing energy~\cite{H_Hiraka2001}. 

In conclusion, we observed an hourglass shape of dispersive magnetic excitation and a peak-dip-hump structure in the energy spectrum 
for underdoped LSCO with $x=0.10$ as is observed in the optimally doped LSCO. 
In the underdoped LSCO, the energy of the $E_{\rm cross}$, where the downward and upward excitations merge with each other, 
is lower than that of the optimally doped LSCO. 
The doping dependence of the $E_{\rm cross}$ is explained as a combined effect of doping-independent spin velocity for the downward excitation
and the doping-dependent incommensurability. 
We discussed the different natures of magnetic excitations in LSCO and YBCO irrespective of the similarity in the hourglass-like dispersion.


We are grateful to K. Hirota and T. J. Sato for experimental support in crystal growth and characterization.
We also thank Y. Bang, C. H. Lee, M. Fujita, and H. Hiraka for stimulating discussions.
The present neutron scattering experiment was performed under the collaboration in neutron science between the High Energy Accelerator Research
Organization and Los Alamos National Laboratory. This work was supported by a Grant-in-Aid for Creative Scientific Research (No.~16GS0417)
,Scientific Research(B) and the Inter-University Research Program on Pulsed-neutron Scattering at Oversea Facilities from MEXT of Japan.


\begin{thebibliography}{25}
\bibitem{note1}
In this paper, we denote the crystallographic indices by the HTT notation, although the crystal structure has an orthorhombic symmetry~($Bmab$) below $T=200$~K.
\bibitem{note2}
In our configuration of measurements, $L$-component of $\bm Q$ changes with increasing energy transfer $\omega$ and also depends on the incident energy.
We use the free-ion magnetic form factor calculated by Freeman and Watson.
\end{thebibliography}

\end{document}